# Systematic study of heavy cluster emission from $^{210-226}$Ra isotopes

K. P. Santhosh[a,*], Sabina Sahadevan, B. Priyanka and M. S. Unnikrishnan

*School of Pure and Applied Physics, Kannur University, Payyanur Campus, Payyanur 670 327, India*

**Abstract**

The half lives for various clusters lying in the cold reaction valleys of $^{210-226}$Ra isotopes are computed using our Coulomb and proximity potential model (CPPM). The computed half lives of $^{4}$He and $^{14}$C clusters from $^{210-226}$Ra isotopes are in good agreement with experimental data. Half lives are also computed using the Universal formula for cluster decay (UNIV) of Poenaru et al., and are found to be in agreement with CPPM values. Our study reveals the role of doubly magic $^{208}$Pb daughter in cluster decay process. Geiger – Nuttall plots for all clusters up to $^{62}$Fe are studied and are found to be linear with different slopes and intercepts. $^{12,14}$C emission from $^{220}$Ra; $^{14}$C emission from $^{222,224}$Ra; $^{14}$C and $^{20}$O emission from $^{226}$Ra are found to be most favourable for measurement and this observation will serve as a guide to the future experiments.

[*]Tel: +919495409757; Fax: +914972806402

*Email address*: drkpsanthosh@gmail.com

## 1. Introduction

The radioactive decay of nuclei emitting particle heavier than alpha particle called cluster radioactivity comes under the wider class of cold decays. The exclusive feature of this phenomenon is the formation of the decay products in the ground or the lowest excited states which contradicts the normal, hot fission, where highly excited fragments are produced. This rare, cold (neutron-less) phenomenon intermediate between alpha decay and spontaneous fission was first predicted by Sandulescu et al. [1] in 1980 on the basis of quantum mechanical fragmentation theory, numerical and analytical superasymmetric fission models, as well as by extending the alpha-decay theory to heavier fragments [2]. The rare nature of this process is due to the fact that cluster emission is marked by several alpha emissions. Experimentally, Rose and Jones [3] first observed such decay in 1984 in the radioactive decay of $^{223}$Ra by the emission of $^{14}$C. The observation of $^{24}$Ne decay from $^{231}$Pa by Tretyakova et al. [4] in Dubna using solid state track detectors (SSTD) was the next important step and the method occurred to be the most effective for cluster radioactivity studies. This was the result which demonstrated that the spontaneous emission of the light nuclei is not limited by a single case of $^{223}$Ra and has a general character in accordance with the predictions [1]. Intense experimental research has led to the detection of about 20 cases of spontaneous emission of clusters ranging from $^{14}$C to $^{34}$Si, from trans-lead nuclei with partial half life ranging from $10^{11}$s to $10^{28}$s [5]. The feature of these emissions is that heavier nuclei will emit heavier fragments in such a way that daughter nuclei are always the doubly magic or nearly doubly magic (i.e. $^{208}$Pb or closely neighboring nuclei).

Many theoretical models have been introduced for explaining cluster radioactivity; these models can be broadly classified as fission-like and alpha-like. One of the alpha-like model is the cluster model. In fission-like model [1, 6-16] the nucleus deforms continuously as it penetrates the nuclear barrier and reaches the scission configuration after running down the Coulomb barrier. In alpha-like model [17-21] the cluster is assumed to be pre-formed in the parent nuclei before it penetrate the nuclear interacting barrier. Since the first experimental observation of cluster radioactivity in 1984, the Analytical Superasymmetric Fission Model have been successfully used to compute half life for alpha and cluster radioactivity in heavy and superheavy nuclides (see the reviews [22, 23] and references therein). Recently Poenaru et al [24] in their letter predicted heavy particle radioactivity (HPR) from elements with Z > 110 leading to doubly magic $^{208}$Pb with shorter half life and larger branching ratio relative to alpha decay. Recently D Ni et al. [25] proposed a unified formula of half-lives for $\alpha$ decay and cluster radioactivity to study the decay of even-even nuclei. Z Ren et al. [26, 27] analysed cluster radioactivity using microscopic density-dependent cluster model with the renormalized M3Y nucleon-nucleon interaction and also reproduced cluster decay half lives using a new formula. The new formula can be considered as a natural extension of Geiger – Nuttall law and Viola-Seaborg formula. Sheng et al. [28] using the effective liquid drop description with the varying mass asymmetry shape and effective inertial coefficient to study cluster decay and improved the result by including the isospin-dependent term in nuclear radius formula.

The study of radium isotopes in the field of cluster decay is prominent because due to their position in the trans-lead region in the nuclear chart the role of the doubly magic $^{208}$Pb and its immediate neighbors as one of the decay fragments comes into play when cluster emission occurs. In fact experimentally $^{14}$C spontaneous emission from radium isotopes is one of the most observed cluster decay modes. Therefore the cluster emission of a few even-even radium isotopes, $^{210-226}$Ra, which leads to the widely studied $^{208}$Pb or neighbouring nuclei as daughters has been studied using the Coulomb and proximity potential model (CPPM) [15, 29, 30, 31] proposed by one of us (KPS) with a view to analyze the spherical shell closures in the daughter lead region. The study will look into all the aspects of alpha and cluster decay from these isotopes beginning with the identification of the most probable clusters from them which will be followed by the study of their decay properties and then finally the results will be analysed to reveal the proton and neutron shell closures in this region. The systematic study on cluster radioactivity from the $^{210\text{-}226}$Ra isotopes has been done for the first time and the $^{210\text{-}226}$Ra isotopes are all alpha instable thereby exhibiting alpha radioactivity. We would like to mention that the present work is an extension of the works presented at various Symposiums on Nuclear Physics [32, 33].

The details of Coulomb and Proximity Potential Model (CPPM) are described in section 2, results and discussions on alpha and cluster decay structure of nuclei are made in section 3 and in section 4 we summarize the entire work.

## 2. The Coulomb and Proximity Potential Model (CPPM)

In Coulomb and proximity potential model (CPPM) the potential energy barrier is taken as the sum of Coulomb potential, proximity potential and centrifugal potential for the touching configuration and for the separated fragments. For the pre-scission (overlap) region, simple power law interpolation as done by Shi and Swiatecki [34] is used. The inclusion of proximity potential reduces the height of the potential barrier, which closely agrees with the experimental result [35]. The proximity potential was first used by Shi and Swiatecki [34] in an empirical manner and has been quite extensively used by Gupta et al [20] in the preformed cluster model (PCM) which is based on pocket formula of Blocki et al. [36] given as:

$$\Phi(\varepsilon) = -\left(\frac{1}{2}\right)(\varepsilon - 2.54)^2 - 0.0852(\varepsilon - 2.54)^3 \text{, for } \varepsilon \leq 1.2511 \qquad (1)$$

$$\Phi(\varepsilon) = -3.437\exp\left(\frac{-\varepsilon}{0.75}\right) \text{, for } \varepsilon \geq 1.2511 \qquad (2)$$

where $\Phi$ is the universal proximity potential. In the present model, another formulation of proximity potential [37] is used as given by Eqs. 6 and 7. In this model cluster formation probability is taken as unity for all clusters irrespective of their masses, so the present model differs from PCM by a factor $P_0$, the cluster formation probability. In the present model assault frequency, ν is calculated for each parent-cluster combination which is associated with vibration energy. But Shi and Swiatecki [38] get ν empirically, unrealistic values $10^{22}$ for even A parent and $10^{20}$ for odd A parent.

The interacting potential barrier for a parent nucleus exhibiting exotic decay is given by,

$$V = \frac{Z_1 Z_2 e^2}{r} + V_p(z) + \frac{\hbar^2 \ell(\ell+1)}{2\mu r^2}, \text{ for } z > 0 \tag{3}$$

Here $Z_1$ and $Z_2$ are the atomic numbers of the daughter and emitted cluster, 'z' is the distance between the near surfaces of the fragments, 'r' is the distance between fragment centers. The term $\ell$ represents the angular momentum, $\mu$ the reduced mass and $V_P$ is the proximity potential. The proximity potential $V_P$ is given by Blocki et al. [36] as,

$$V_p(z) = 4\pi\gamma b\left[\frac{C_1 C_2}{(C_1 + C_2)}\right]\Phi\left(\frac{z}{b}\right), \tag{4}$$

with the nuclear surface tension coefficient,

$$\gamma = 0.9517[1 - 1.7826(N-Z)^2 / A^2] \text{ MeV/fm}^2 \tag{5}$$

where N, Z and A represent neutron, proton and mass number of parent respectively, $\Phi$ represents the universal the proximity potential [37] given as

$$\Phi(\varepsilon) = -4.41 e^{-\varepsilon/0.7176}, \text{ for } \varepsilon \geq 1.9475 \tag{6}$$

$$\Phi(\varepsilon) = -1.7817 + 0.9270\varepsilon + 0.0169\varepsilon^2 - 0.05148\varepsilon^3, \text{ for } 0 \leq \varepsilon \leq 1.9475 \tag{7}$$

With ε = z/b, where the width (diffuseness) of the nuclear surface b ≈ 1 and Süsmann central radii $C_i$ of fragments related to sharp radii $R_i$ as,

$$C_i = R_i - \left(\frac{b^2}{R_i}\right) \tag{8}$$

For $R_i$ we use semi empirical formula in terms of mass number $A_i$ as [36],

$$R_i = 1.28A_i^{1/3} - 0.76 + 0.8A_i^{-1/3} \tag{9}$$

The potential for the internal part (overlap region) of the barrier is given as,

$$V = a_0(L - L_0)^n, \text{ for } z < 0 \tag{10}$$

Here $L = z + 2C_1 + 2C_2$ and $L_0 = 2C$, the diameter of the parent nuclei. The constants $a_0$ and $n$ are determined by the smooth matching of the two potentials at the touching point.

Using one dimensional WKB approximation, the barrier penetrability P is given as,

$$\mathrm{P} = \exp\left\{ -\frac{2}{\hbar}\int_a^b \sqrt{2\mu(V - Q)}dz \right\} \tag{11}$$

Here the mass parameter is replaced by $\mu = mA_1A_2/A$, where m is the nucleon mass and $A_1$, $A_2$ are the mass numbers of daughter and emitted cluster respectively. The turning points "$a$" and "b" are determined from the equation $V(a) = V(b) = Q$. The above integral can be evaluated numerically or analytically, and the half life time is given by

$$T_{1/2} = \left(\frac{\ln 2}{\lambda}\right) = \left(\frac{\ln 2}{\upsilon P}\right) \tag{12}$$

where, $\upsilon = \left(\frac{\omega}{2\pi}\right) = \left(\frac{2E_v}{h}\right)$ represent the number of assaults on the barrier per second and λ the decay constant. $E_v$, the empirical vibration energy is given as [7],

$$E_v = Q\left\{0.056 + 0.039\exp\left[\frac{(4 - A_2)}{2.5}\right]\right\}, \text{ for } A_2 \geq 4 \tag{13}$$

In the classical method, the cluster is assumed to move back and forth in the nucleus and the usual way of determining the assault frequency is through the expression given by $\nu = velocity/(2R)$, where $R$ is the radius of the parent nuclei. But the cluster has wave properties; therefore a quantum mechanical treatment is more accurate. Thus, assuming that the cluster vibrates in a harmonic oscillator potential with a frequency $\omega$, which depends on the vibration energy $E_v$, we can identify this frequency as the assault frequency $v$ given in eqns. (12)-(13).

## 3. Results and discussions

The decay studies starts with the identification of the probable clusters from the $^{210-226}$Ra isotopes through the cold valley plots. The concept of cold valley was introduced in relation to the structure of minima in the so- called driving potential. The driving potential is defined as the difference between the interaction potential V and the decay energy Q of the reaction. The decay energy Q is given by,

$$Q = M(A,Z) - M(A_1,Z_1) - M(A_2,Z_2), \tag{14}$$

where $M(A,Z)$, $M(A_1,Z_1)$, $M(A_2,Z_2)$ are the atomic masses of parent, daughter and emitted cluster respectively. The possibility to have a cluster decay process is that the decay energy of the reaction (Q-value) must be greater than zero. The barrier penetrability is very sensitive to the Q value and is computed using experimental mass tables of Audi et.al. [39] wherever possible. In the cases where the experimental mass excess values are not available we have calculated the Q value using the mass tables of KTUY [40]. So full shell effects are contained in our model that comes from experimental and/or calculated mass excesses.

The driving potential $(V - Q)$ for a particular parent is calculated for all possible cluster-daughter combinations as a function of mass and charge asymmetries, $\eta = \frac{A_1 - A_2}{A_1 + A_2}$ and $\eta_Z = \frac{Z_1 - Z_2}{Z_1 + Z_2}$, at the touching configuration. For every fixed mass pair ($A_1$, $A_2$) a pair of charges is singled out for which driving potential is minimum.

Figures 1-3 give the plots of driving potential versus $A_2$, the mass number of one of the fragments (cold valley plots) for $^{210-226}$Ra isotopes. The occurrence of the mass – asymmetry valleys in these figures is due to the shell effects. The minima in driving potential

represent the most probable decay which is due to the shell closure of one or both fragments. The figures reveal that the clusters most probable for emission from all of them are the typical clusters of $^{4}He$, $^{8, 10}Be$, $^{12, 14}C$, $^{18, 20}O$, $^{22, 24}Ne$, $^{26, 28}Mg$ etc. In addition to these usual clusters some other deep valleys can also be observed in the graphs of all the parents. One valley which appears in the plots for all the parent isotopes is the Ca valley with $^{46-52}Ca$ nuclei as the probable clusters. These valleys signify the role of the Z = 20, N = 28 magicities in the Ca nuclei. Another valley which appears in all the graphs is the one at the Sr clusters which are of course due to the Z = 50 spherical shell closure in the corresponding Sn daughter fragments. In between these two valleys a third valley comes up at the Ni cluster for $^{210}Ra$ but it gets shifted to Zn cluster for $^{212}Ra$ until it merges with the Sr valley as a dip at $^{92}Kr$ in the cold valley graph for $^{226}Ra$. This happens so because this deep valley is due to the N = 82 magic number in the daughter nuclei. Thus the well known spherical shell closures in this region of the nuclear chart are getting reflected in the cold valleys for the radium isotopes.

The Hartee-Fock method [41] is able to reproduce experimental Q value even though it does not take care of the shell effect. In the present paper we have not calculated shell effects explicitly as in microscopic-macroscopic model, but the shell effects are incorporated in our model that comes through the experimental Q value. If the experimental Q value (not the shell effect) were the cause for the staggering in heights of driving potential, we would like to mention that experimental Q values are taken for the computation of driving potential, so all points in the driving potential plots (Figs 1-3) will show a staggering which results in the disappearance of cold valley. This emphasizes the fact that staggering in height is due to the shell effect (also see Ref. [42, 43]).

After identification of the most probable clusters from the driving potential versus $A_2$ graphs the decay characteristics are computed within CPPM. The Q values and half lives for the emission of various clusters from the $^{210\text{-}226}$Ra parent isotopes are tabulated in Table 1. The half life calculations are also done using universal formula for the cluster decay (UNIV) of Poenaru et al. [44] given as

$$\log_{10} T_{1/2}(s) = -\log_{10} P - \log_{10} S + [\log_{10}(\ln 2) - \log_{10} \nu] \qquad (15)$$

where $\nu$ is a constant and $S$ is the preformation probability of the cluster at the nuclear surface which depends only on the mass number of the emitted cluster. The decimal logarithm of the preformation factor is given as

$$\log_{10} S = -0.598(A_e - 1) \qquad (16)$$

In the case of an even-even nucleus, the additive constant is denoted as

$$c_{ee} = [-\log_{10} \nu + \log_{10}(\ln 2)] = -22.16917 \qquad (17)$$

The penetrability of an external Coulomb barrier, having the first turning point as the separation distance at the touching configuration $R_a = R_t = R_d + R_e$ and the second turning point defined by $e^2 Z_d Z_e / R_b = Q$, may be found analytically as

$$-\log_{10} P = 0.22873(\mu_A Z_D Z_e R_b)^{1/2} \times [\arccos(\sqrt{r} - \sqrt{r(1-r)})] \qquad (18)$$

where $r = R_t / R_b$, $R_t = 1.2249(A_d^{1/3} + A_e^{1/3})$, and $R_b = 1.43998 Z_d Z_e / Q$.

In the Table 1, the half life calculations done using our formalism CPPM is given in column 5 and the calculations done using the universal formula (UNIV) are given in column 6. The experimental $T_{1/2}$ values for alpha decay are taken from [45] and that of $^{14}$C are obtained from Ref [46, 47, 48] and are given in column 7. It is clear from these tables that our calculated values matches well with the experimental values and the values calculated using the universal formula (UNIV). In Analytical Superasymmetric Fission Model, Poenaru et al. used a second order polynomial for the overlap (pre-scission) region but in present model a power law interpolation is taken which are not quadratic, and the nuclear part in the potential for post-scission region is replaced by proximity potential in our model. So a deviation in computed half life values are expected due to the difference in potential used in both models, but the calculation shows CPPM values match with UNIV values.

Figure 4 gives the $\log_{10}(T_{1/2})$ versus neutron number of parent graph for the $^{4}$He, $^{12,14}$C, $^{16,18,20}$O, $^{22,24}$Ne decays. The experimental $\log_{10}(T_{1/2})$ values for $^{4}$He and $^{14}$C clusters are compared with the CPPM values and they are in good agreement as can be seen from Figure 4 and from Table 1. For e.g. in the case of α-decay from $^{214}$Ra $\{\log_{10}(T_{1/2})\}_{calc.} = 0.48$ and $\{\log_{10}(T_{1/2})\}_{expt.} = 0.39$ and that of $^{14}$C decay from $^{222}$Ra $\{\log_{10}(T_{1/2})\}_{calc.} = 11.02$ and $\{\log_{10}(T_{1/2})\}_{expt.} = 11.1$. In Figure 4 a slight rise in $T_{1/2}$ values is got for all the clusters at N = 126 signifying the well established spherical shell closure there. The graph is also showing dips at certain neutron numbers thereby denoting the shell closure in the daughter nucleus there. The first dip is got at $^{216}$Ra parent (N = 128) for the alpha decay and this is due to the N = 126 magicity in the $^{212}$Rn daughter. Then there are dips at N = 132, 134 and 136 ($^{222, 224, 226}$Ra) for $^{12}$C, $^{14}$C and $^{16}$C emissions respectively. All the three decrease in half lives are due to the double magicity (Z = 82, N =126) of the well known $^{208}$Pb daughters. Two other dips are observed at the $^{224}$Ra (N = 136) parent for the decay of $^{18}$O and $^{24}$Ne emissions respectively. The first one is due to the N = 126 spherical closure in the $^{206}$Hg daughter and the second one is due to Z ≈ 76 major (deformed) closed shell in the $^{200}$Pt daughter nucleus.

Figure 5 displays similar plot for $^{26, 28}$Mg, $^{30, 32, 34}$Si and $^{36, 38, 40}$S cluster emissions. Here too there is a rise in $T_{1/2}$ values at N = 126 which is expected. Dips are seen at N = 128 ($^{216}$Ra) for $^{36}$S decay at N = 130 for $^{32}$Si and $^{38}$S decays respectively. The dip for $^{36}$S emission is due to the Z = 72 major (deformed) shell closure in $^{180}$Hf daughter while the dips at N = 130 ($^{220}$Ra) for $^{32}$Si and $^{38}$S decays are due to the Z = 74, 72 major (deformed) shell closures in $^{186}$W and $^{180}$Hf daughters respectively. Again a dip can be seen at $^{222}$Ra parent (N = 134) for $^{40}$S decay and this is also due to the Z = 72 deformed shell closure in the $^{182}$Hf daughter fragment. The Z = 76, 74, 72 are major (deformed) shell closures which are predicted by Santhosh et al. [49] and references therein. Therefore the study on Ra isotopes is exhibiting all the known spherical shell closures in the relevant part of the nuclear chart and together with it some deformed shell closures are also coming up from the half lives graphs.

Figure 6 gives the Geiger–Nuttall plots for $\log_{10}(T_{1/2})$ values vs $Q^{-1/2}$ for $^{4}$He, $^{8}$Be, $^{10}$Be, $^{12}$C, $^{14}$C, $^{18}$O, $^{24}$Ne, $^{28}$Mg, $^{34}$Si, $^{38}$S, $^{44}$Ar, $^{48}$Ca, $^{54}$Ti, $^{58}$Cr, and $^{62}$Fe emissions from $^{210-226}$Ra isotopes. Geiger–Nuttall plots for all clusters are found to be linear with different slopes and intercepts. We would like to point out that Geiger Nuttall law is for pure Coulomb potential, but from our present study it is found that inclusion of proximity potential will not produce any deviation to the linear nature of these Geiger Nuttall plots. We would also like to mention that the presence of proximity potential (nuclear structure effect) and shell effect (through Q value) are evident from the observed variation in the slope and intercept of Geiger Nuttall plots for different clusters from Ra isotopes.

The recent studies using Hartee-Fock or mic-mac method [50, 51, 52] were able to reproduce cluster decay half lives very well but it is interesting to see the balance between the difficulty and complexity of Hartee-Fock or mic-mac method and the precision and prediction of our simple model. Using presently available techniques the longest measured half life is of the order of $10^{30}$s and the lowest measurable branching ratio is almost $10^{-19}$. We have also computed branching ratio of cluster decay with respect to alpha emission, for all clusters emitting from $^{210-226}$Ra isotopes. From the half life and branching ratio consideration, it is found that $^{12,14}$C emission from $^{220}$Ra; $^{14}$C emission from $^{222,224}$Ra; $^{14}$C and $^{20}$O emission from $^{226}$Ra are the most favourable for measurement. We hope this observation will serve as a guide to the future experiments.

## 4. Conclusion

The cold valley plots for the $^{210-226}$Ra isotopes are analyzed to determine the various clusters possible for emission from each of the Ra isotopes. The half life times and other characteristics for the possible cluster emissions are computed and tabulated. The $\log_{10}(T_{1/2})$ versus neutron number of parent graphs for the alpha and cluster decays are plotted from which the well established spherical shell closures at Z = 82 and N = 126 are clearly coming up as elevations and dips in the graphs. In addition to these spherical shell closures, major (deformed) shell closures at Z = 76, 74, 72 can also be observed from the half lives graphs. $^{12,14}C$ emission from $^{220}Ra$; $^{14}C$ emission from $^{222,224}Ra$; $^{14}C$ and $^{20}O$ emission from $^{226}Ra$ are found to be most favourable for measurement and this observation will serve as a guide to the future experiments.

**References**

[1] A Sandulescu, D N Poenaru and W Greiner, Sov. J. Part. Nucl. **II** 528 (1980)

[2] A Sandulescu, D N Poenaru, W Greiner and J H Hamilton, Phys. Rev. Lett. **54** 490 (1985)

[3] H J Rose and G A Jones, Nature **307** 245 (1984)

[4] A Sandulescu, Yu S Zamyatnin, I A Lebedev, V F Myasoedov, S P Tretyakova and D Hashegan, JINR Rapid Commun. **5** 84 (1984)

[5] R Bonetti and A Guglielmetti, Rom. Rep. Phys. **59** 301 (2007)

[6] D N Poenaru, M Ivascu, A Sandulescu and W Greiner, J. Phys. G: Nucl Part. Phys. **10** L183 (1984)

[7] D N Poenaru, M Ivascu, A Sandulescu and W Greiner, Phys. Rev. C **32** 572 (1985)

[8] D N Poenaru, W Greiner, A Sandulescu and M Ivascu, Phys. Rev. C **32** 2198 (1985)

[9] W Greiner, M Ivascu, D N Poenaru and A Sandulescu, Z. Phys. A **320** 347 (1985)

[10] Y J Shi and W J Swiatecki, Phys. Rev. Lett. **54** 300 (1985)

[11] G A Pik-Pichak, Sov. J. Nucl. Phys. **44** 923 (1986)

[12] G Shanmugam and B Kamalaharan, Phys. Rev. C **38** 1377 (1988)

[13] B Buck and A C Merchant, J. Phys. G: Nucl. Part. Phys. **15** 615 (1989)

[14] K P Santhosh and A Joseph, Ind. J. of Pure & App. Phys. **42** 806 (2004)

[15] K P Santhosh and A Joseph, Pramana J. Phys **55** 375 (2000)

[16] G Royer, R K Gupta and V Yu Denisov, Nucl. Phys. A **632** 275 (1998)

[17] S Landowne and C H Dasso, Phys. Rev. C **33**, 387 (1986)

[18] M Iriondo, D Jerrestan and R J Liotta, Nucl. Phys. A **454** 252 (1986)

[19] R Blendoaske, T Fliessbach and H Walliser, Nucl. Phys A **464** 75 (1987)

[20] S S Malik and R K Gupta, Phys. Rev. C **39** 1992 (1989)

[21] S Kumar and R K Gupta, Phys. Rev. C **55** 218 (1997)

[22] D N Poenaru and W Greiner in Clusters in Nuclei. Lecture Notes in Physics 1, edited by C Beck (Springer, Berlin, 2010) Vol. **818** p.1

[23] H J Krappe and K Pomorski in Theory of Nuclear Fission: A Textbook. Lecture Notes in Physics (Springer, Berlin, 2010) Vol. **838** p.233

[24] D N Poenaru, R A Gherghescu and W Greiner, Phys. Rev. Lett. **107** 062503 (2011)

[25] D Ni, R Z en, T Dong and C Xu, Phys. Rev. C **78** 044310 (2008)

[26] Z Ren, C Xu and Z Wang, Phys. Rev. C **70** 034304 (2004)

[27] D Ni and Z Ren, Phys. Rev. C **82** 024311(2010)

[28] Z Sheng, D Ni and Z Ren, J. Phys. G: Nucl. Part. Phys. **38** 055103 (2011)

[29] K P Santhosh and A Joseph, Proc. Int. Nat. Symp. Nucl. Phys. (India) **43B** 296 (2000)

[30] K P Santhosh and A Joseph, Pramana J. Phys. **58** 611 (2002)

[31] K P Santhosh, R K Biju, and A Joseph, J. Phys. G: Nucl. Part. Phys. **35** 085102 (2008)

[32] K P Santhosh, T Ambily, S Sabina, R K Biju and A Joseph, Proc. Nat. Symp. Nucl. Phys. (India) **1** 33 (2009)

[33] K P Santhosh, S Sabina, B Priyanka, M S Unnikrishnan, G J Jayesh and R K Biju, Proc. DAE Symp. Nucl. Phys. (India) **56** 298 (2011)

[34] Y J Shi and W J Swiatecki, Nucl. Phys. A **438** 450 (1985)

[35] K P Santhosh and A Joseph, Pramana J. Phys. **62** 957 (2004)

[36] J Blocki, J Randrup, W J Swiatecki and C F Tsang, Ann. Phys. NY **105** 427 (1977)

[37] J Blocki and W J Swiatecki, Ann. Phys. NY **132** 53 (1981)

[38] Y J Shi and W J Swiatecki, Nucl. Phys. A **464** 205 (1987)

[39] G Audi, A H Wapstra and C Thivault, Nucl. Phys. A **729** 337 (2003)

[40] H Koura, T Tachibana, M Uno and M Yamada, Prog. Theor. Phys. **113** 305 (2005)

[41] M Bhuyan, S K Patra, and R K Gupta, Phys. Rev. C **84** 014317 (2011)

[42] R K Gupta and W Greiner, Int. J. Mod. Phys. E **3** 335 (1994)

[43] R K Gupta, Pramana J. Phys. **57** 481 (2001)

[44] D N Poenaru, R A Gherghescu and W Greiner, Phys. Rev. C **83** 014601 (2011)

[45] NuDat2.5, http://www.nndc.bnl.gov

[46] P B Price, J D Stevenson, S W Barwick and H L Ravn, Phys. Rev. Lett. **54** 297 (1985)

[47] E Hourani, L Rosier, G Berrier-Ronsin, Elayi, A C Mueller, G Rappenecker, G Rotbard, G Renou, A Libe and L Stab, Phys. Rev. C **44** 1424 (1991)

[48] E Hourani, M Hussonnois, L Stab, L Brillard, S Gales and J P Schapira, Phys. Lett. B **160** 375 (1985)

[49] K P Santhosh and S Sabina, Nucl. Phys. A **847** 42 (2010)

[50] B B Singh, S K Patra and R K Gupta, Phys. Rev. C **82** 014607 (2010).

[51] M Warda and L M Robledo, Phys. Rev. C **84**, 044608 (2011)

[52] B B Singh, M Bhuyan, S K Patra and R K Gupta J. Phys. G: Nucl. Part. Phys. **39** 025101 (2012)

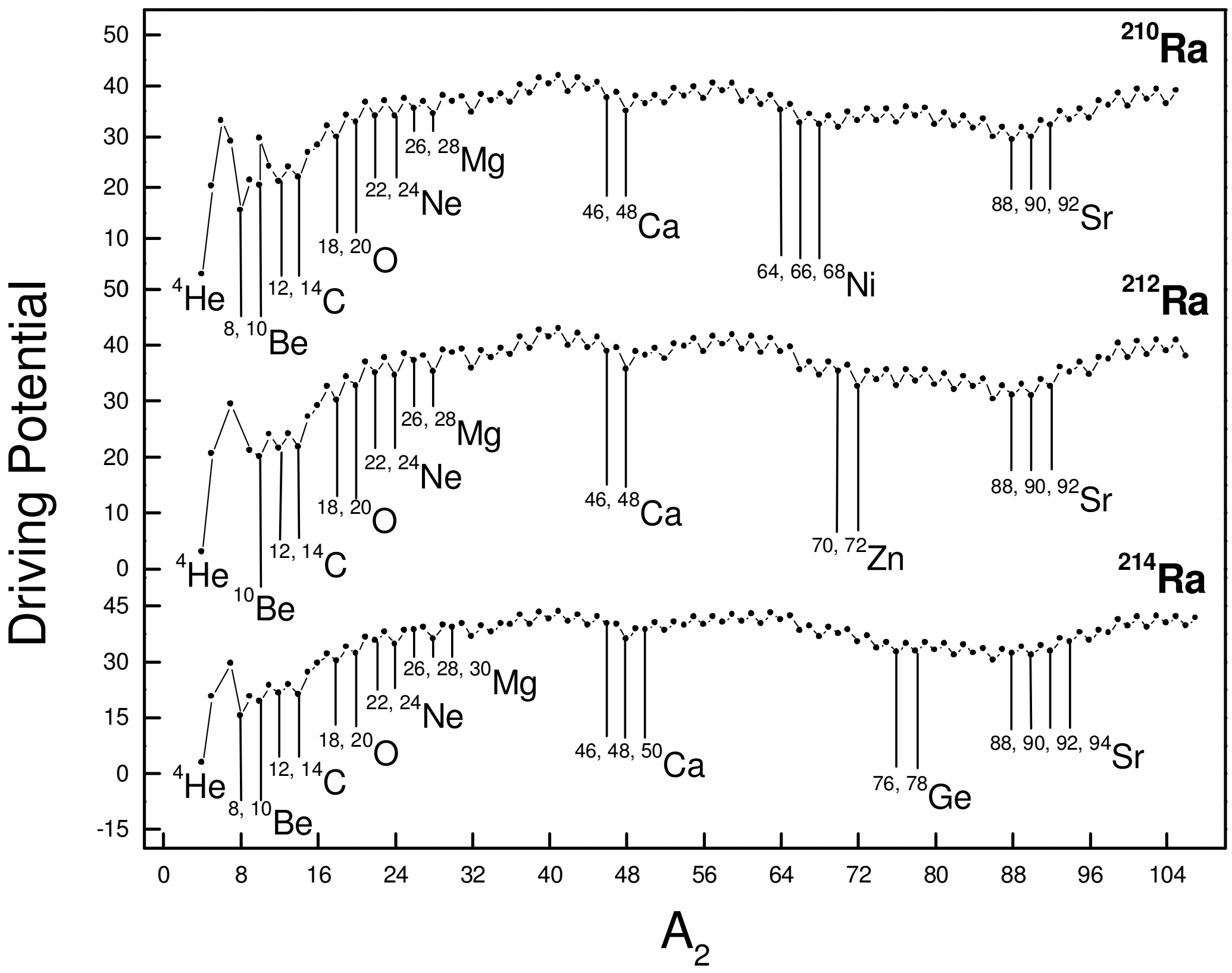

Fig. 1. The driving potentials as a function of the mass number of one of the fragments for $^{210, 212, 214}$Ra isotopes. The calculations are made for touching configuration, $r = C_1 + C_2$.

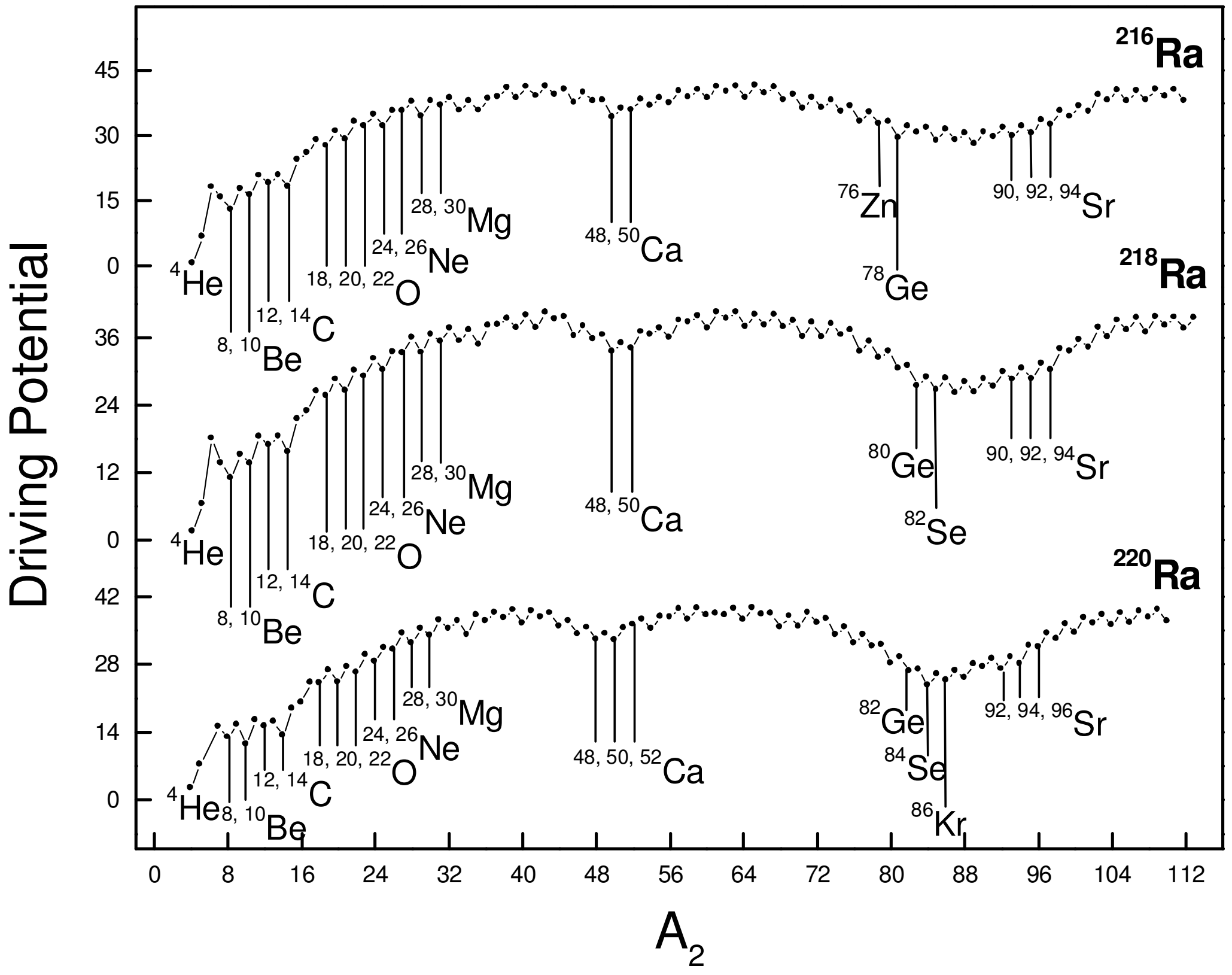

Fig. 2. The driving potentials as a function of the mass number of one of the fragments for $^{216, 218, 220}$Ra isotopes. The calculations are made for touching configuration, $r = C_1 + C_2$.

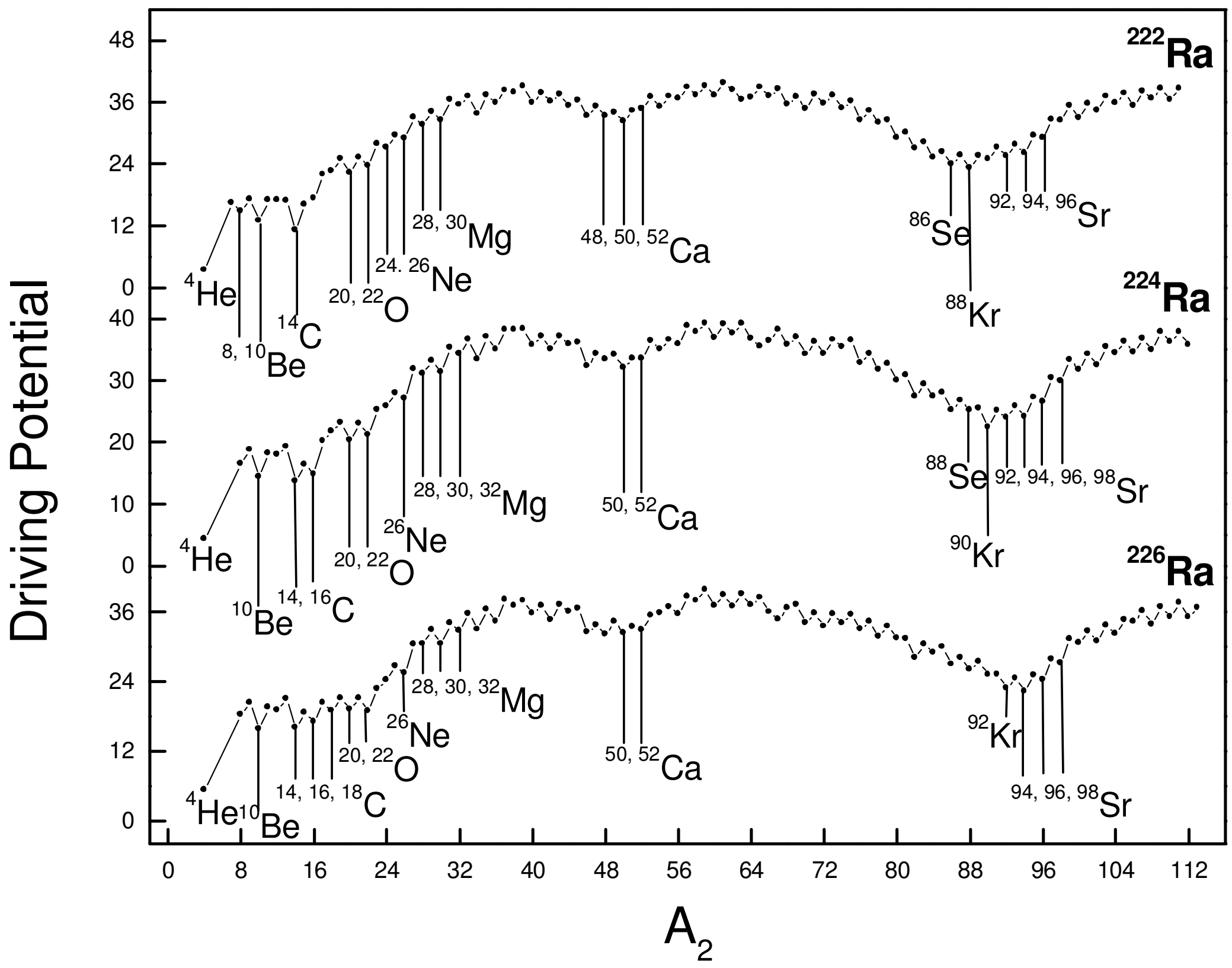

Fig. 3. The driving potentials as a function of the mass number of one of the fragments for $^{222, 224, 226}Ra$ isotopes. The calculations are made for touching configuration, $r = C_1 + C_2$.

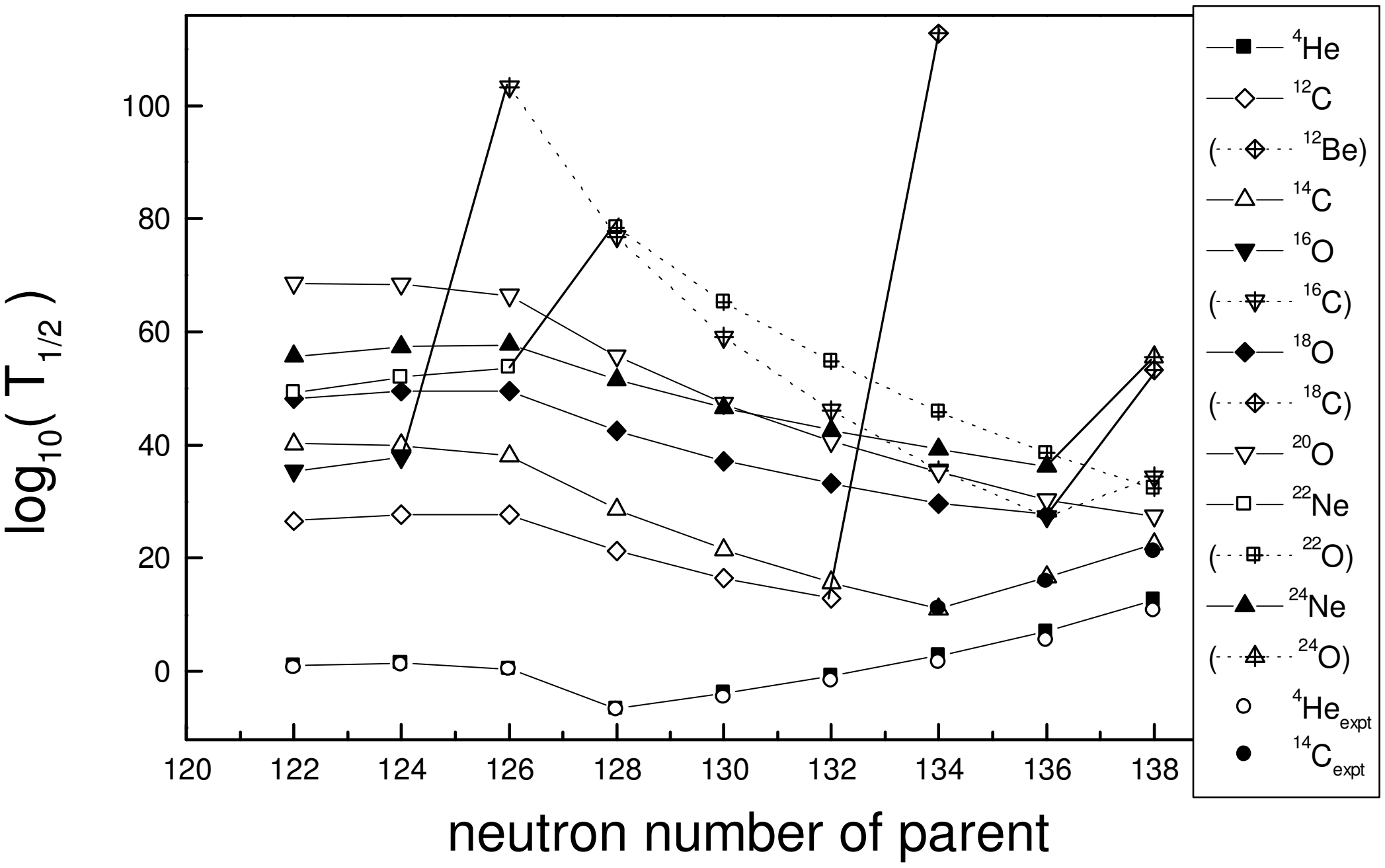

Fig. 4. Plot of the computed $\log_{10}(T_{1/2})$ values vs. neutron number of parent for $^{4}$He, $^{12,\,14}$C, $^{16,\,18,\,20}$O and $^{22,\,24}$Ne clusters from Ra isotopes

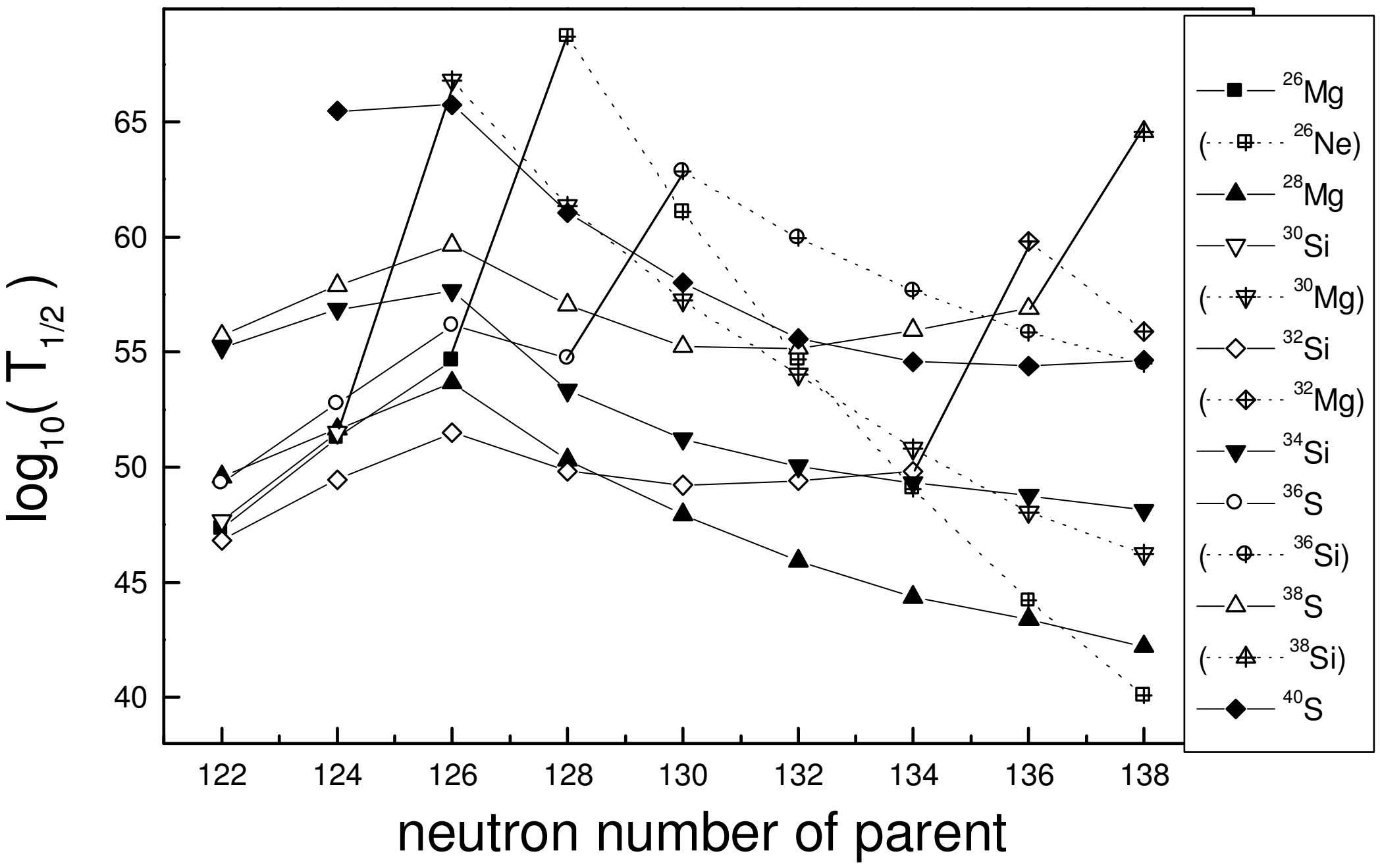

Fig. 5. Plot of the computed $\log_{10}(T_{1/2})$ values vs. neutron number of parent for $^{26,28}$Mg, $^{30,32,34}$Si and $^{36,38,40}$S clusters from Ra isotopes.

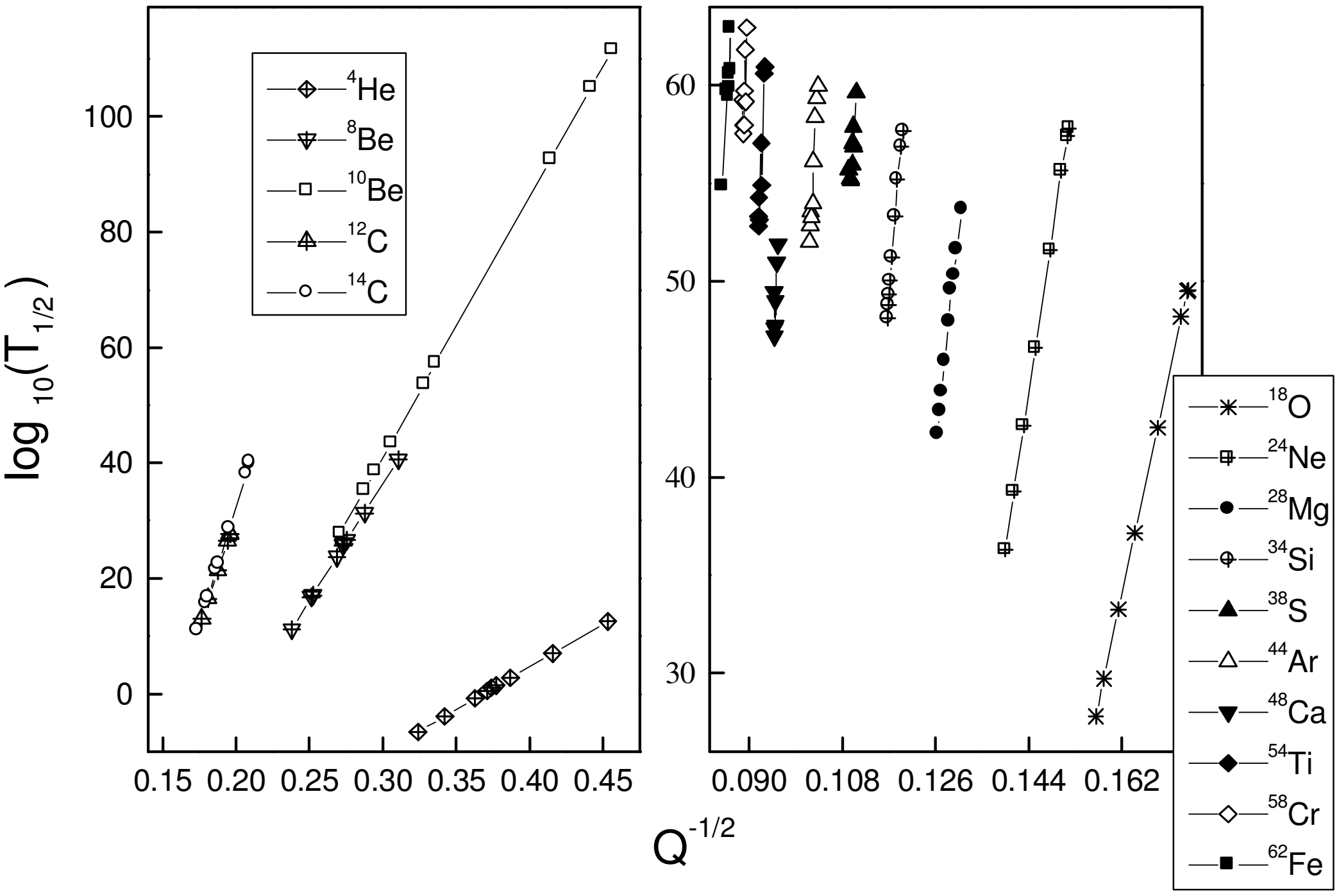

Fig. 6. Geiger – Nuttall plot for $\log_{10}(T_{1/2})$ values vs $Q^{-1/2}$ for various clusters from $^{210\text{-}226}Ra$ isotopes.

Table 1. Computed Q value and logarithm of half life times of $^{210-226}$Ra decaying by the emission of various probable clusters. $T_{1/2}$ is in seconds.

| Parent nuclei | Emitted cluster | Daughter nuclei | Q value (MeV) | $\log_{10}(T_{1/2})$ | | | |
|---|---|---|---|---|---|---|---|
| | | | | CPPM | UNIV | Expt. | Ref. |
| $^{210}$Ra | $^{4}$He | $^{206}$Rn | 7.152 | 1.02 | 0.21 | 0.55 | [45] |
| $^{210}$Ra | $^{8}$Be | $^{202}$Po | 13.443 | 25.75 | 22.78 | | |
| $^{210}$Ra | $^{12}$C | $^{198}$Pb | 26.511 | 26.58 | 22.47 | | |
| $^{210}$Ra | $^{16}$O | $^{194}$Hg | 37.391 | 35.43 | 28.83 | | |
| $^{212}$Ra | $^{4}$He | $^{208}$Rn | 7.032 | 1.46 | 0.61 | 1.04 | [45] |
| $^{212}$Ra | $^{8}$Be | $^{204}$Po | 13.201 | 26.72 | 23.64 | | |
| $^{212}$Ra | $^{12}$C | $^{200}$Pb | 26.052 | 27.72 | 23.46 | | |
| $^{212}$Ra | $^{14}$C | $^{198}$Pb | 22.839 | 39.94 | 36.22 | | |
| $^{212}$Ra | $^{16}$O | $^{196}$Hg | 36.373 | 37.79 | 30.82 | | |
| $^{214}$Ra | $^{4}$He | $^{210}$Rn | 7.274 | 0.48 | -0.29 | 0.39 | [45] |
| $^{214}$Ra | $^{8}$Be | $^{206}$Po | 13.341 | 26.02 | 23.03 | | |
| $^{214}$Ra | $^{12}$C | $^{202}$Pb | 26.035 | 27.65 | 23.41 | | |
| $^{214}$Ra | $^{14}$C | $^{200}$Pb | 23.324 | 38.10 | 34.59 | | |
| $^{216}$Ra | $^{4}$He | $^{212}$Rn | 9.526 | -6.58 | -6.64 | -6.74 | [45] |
| $^{216}$Ra | $^{8}$Be | $^{208}$Po | 15.819 | 16.79 | 14.64 | | |
| $^{216}$Ra | $^{12}$C | $^{204}$Pb | 28.401 | 21.35 | 18.05 | | |
| $^{216}$Ra | $^{14}$C | $^{202}$Pb | 26.205 | 28.68 | 26.42 | | |
| $^{218}$Ra | $^{4}$He | $^{214}$Rn | 8.546 | -3.90 | -4.27 | -4.59 | [45] |
| $^{218}$Ra | $^{8}$Be | $^{210}$Po | 17.662 | 11.16 | 9.66 | | |
| $^{218}$Ra | $^{12}$C | $^{206}$Pb | 30.436 | 16.47 | 14.01 | | |
| $^{218}$Ra | $^{14}$C | $^{204}$Pb | 28.741 | 21.49 | 20.37 | | |
| $^{218}$Ra | $^{18}$O | $^{200}$Hg | 36.937 | 37.16 | 32.29 | | |
| $^{220}$Ra | $^{4}$He | $^{216}$Rn | 7.592 | -0.80 | -1.48 | -1.74 | [45] |
| $^{220}$Ra | $^{8}$Be | $^{212}$Po | 15.701 | 17.04 | 14.86 | | |
| $^{220}$Ra | $^{10}$Be | $^{210}$Po | 13.619 | 27.78 | 26.87 | | |
| $^{220}$Ra | $^{12}$C | $^{208}$Pb | 32.022 | 12.93 | 11.15 | | |
| $^{220}$Ra | $^{14}$C | $^{206}$Pb | 31.039 | 15.67 | 15.63 | | |
| $^{220}$Ra | $^{18}$O | $^{202}$Hg | 38.400 | 33.26 | 29.12 | | |

Table 1 (Continued)

| Parent nuclei | Emitted cluster | Daughter nuclei | Q value (MeV) | $\log_{10}(T_{1/2})$ CPPM | UNIV | Expt. | Ref. |
|---|---|---|---|---|---|---|---|
| $^{222}Ra$ | $^{4}He$ | $^{218}Rn$ | 6.679 | 2.76 | 1.78 | 1.58 | [45] |
| $^{222}Ra$ | $^{8}Be$ | $^{214}Po$ | 13.849 | 23.71 | 20.86 | | |
| $^{222}Ra$ | $^{14}C$ | $^{208}Pb$ | 33.050 | 11.02 | 11.95 | 11.01 | [46] |
| $^{222}Ra$ | $^{18}O$ | $^{204}Hg$ | 39.793 | 29.72 | 26.28 | | |
| $^{222}Ra$ | $^{20}O$ | $^{202}Hg$ | 37.869 | 35.29 | 32.83 | | |
| $^{224}Ra$ | $^{4}He$ | $^{220}Rn$ | 5.789 | 7.01 | 5.74 | 5.49 | [45] |
| $^{224}Ra$ | $^{8}Be$ | $^{216}Po$ | 12.102 | 31.28 | 27.87 | | |
| $^{224}Ra$ | $^{14}C$ | $^{210}Pb$ | 30.536 | 16.67 | 16.41 | 15.68 | [47] |
| $^{224}Ra$ | $^{16}C$ | $^{208}Pb$ | 26.882 | 27.39 | 27.51 | | |
| $^{224}Ra$ | $^{18}O$ | $^{206}Hg$ | 40.555 | 27.79 | 24.75 | | |
| $^{224}Ra$ | $^{20}O$ | $^{204}Hg$ | 39.720 | 30.34 | 28.86 | | |
| $^{226}Ra$ | $^{4}He$ | $^{222}Rn$ | 4.871 | 12.56 | 10.99 | 10.70 | [45] |
| $^{226}Ra$ | $^{14}C$ | $^{212}Pb$ | 28.197 | 22.53 | 21.20 | 21.19 | [48] |
| $^{226}Ra$ | $^{16}C$ | $^{210}Pb$ | 24.703 | 34.38 | 33.45 | | |
| $^{226}Ra$ | $^{20}O$ | $^{206}Hg$ | 40.818 | 27.48 | 26.62 | | |
| $^{226}Ra$ | $^{22}O$ | $^{204}Hg$ | 39.079 | 32.32 | 32.53 | | |